\documentstyle[prl,aps,graphicx,multicol,amssymb]{revtex}

\begin{document}

\draft

\title{Landau's quasi-particle mapping:
Fermi liquid approach and Luttinger liquid behavior}

\author{Caspar P.~Heidbrink and G\"otz S.~Uhrig}
\address{Institut f\"{u}r Theoretische Physik,
Universit\"{a}t zu K\"{o}ln, Z\"{u}lpicher Stra{\ss}e 77,
50937 K\"{o}ln, Germany}

\date{\today}

\maketitle
\begin{abstract}
A continuous unitary transformation is introduced which realizes
Landau's mapping of the elementary excitations (quasi-particles) 
of an interacting Fermi liquid system
to those of the system without interaction.
The conservation of the number of quasi-particles is important.
The transformation is performed numerically for a one-dimensional
system, i.e.\ the worst case for a Fermi liquid approach.
Yet evidence for Luttinger liquid behavior is found. Such an approach may
open a route to a unified description of Fermi and Luttinger liquids
on all energy scales.
\end{abstract}



\begin{multicols}{2}
The behavior of strongly correlated electron systems is a
fundamental issue with  many intriguing but unresolved questions
\cite{orens00,sachd00,ander00}. Two antithetic  concepts
for non-symmetry-broken phases govern the discussion: the Fermi
liquid (FL) \cite{nozie97} and the Luttinger liquid (LL) \cite{voit95}. The
FL, well-established in three-dimensional (3D) systems, is
based on Landau's conjecture that the low-lying excitations of
interacting fermion systems can be  connected continuously to
those of the non-interacting Fermi gas: there is  a smooth mapping
between the quasi-particles of the interacting and of the
non-interacting system. The LL, well-established in
one dimension (1D), is based on the observation that collective
particle-hole excitations exhaust all degrees of freedom. These
collective excitations are bosonic and thus completely different
from the non-interacting fermionic excitations. In two dimensions (2D)
-- relevant for the understanding for high-temperature 
superconductivity
\cite{orens00,sachd00,ander00} -- the situation is marginal
\cite{metzn98} due to the
occurrence of logarithmic factors. Here we
report a non-perturbative prescription based on quasi-particle
conservation to construct Landau's mapping. 
Evidence is found that LL behavior is retrieved
with our method.

FL theory is most successful as
phenomenological theory based on the energy functional
\cite{nozie97}
\begin{equation}
\label{energfunc}
 E= \sum_k \varepsilon(k) \delta n_k + \sum_{k,q} f(k,q)
\delta n_k \delta n_q
\end{equation}
where $k, q$ are momenta close to the Fermi surface, $\delta n_k$
the density of quasi-particles -- particles above the Fermi sea,
holes below, $\varepsilon(k)$ the kinetic energy and $f(k,q)$ is
Landau's interaction function.  
We highlight that
Eq.~\ref{energfunc} implies the conservation of the number of
quasi-particles. Microscopic derivations of Eq.~\ref{energfunc}
are based on perturbation theory \cite{nozie97}. Ordinary
electrons or holes are good quasi-particles on the Fermi surface
$k=k_{\rm F}$ due to the restricted scattering phase space. The
quasi-particle weight $0<Z\le1$ quantifies to which
extent, i.e.\ to which fraction,
 real electrons or holes overlap with  true quasi-particles. Moreover,
the renormalization group (RG) has been used in the last decade to give
FL theory a firm basis
\cite{feldm95,shank94,chito95,dupui98}. Additionally,
RG approaches \cite{salmh98a,salmh01} are recently
 used to understand the 2D Hubbard models in
the context of high-$T_c$ superconductivity
\cite{zanch97,halbo00a,honer01}. All these approaches are
conceptually based on diagrammatic perturbation theory in the
interaction strength. They do not employ or construct Landau's
mapping except on the Fermi surface. Quasi-particle conservation
does not appear.

We use continuous unitary transformations (CUT)
\cite{wegne94,nicod95,kabel97} to treat interacting Fermi systems
\begin{eqnarray}
H &=& E_0+\sum_k \varepsilon(k) :n_k: + \nonumber\\
\label{hamilton}
&& \sum_{K,p,q} \Gamma(K,p,q) :c^\dagger_{K+p}
c^\dagger_{K-p}c^{\phantom\dagger}_{K-q} c^{\phantom\dagger}_{K+q}:
\end{eqnarray}
where $: \ :$ stands for normal ordering (in particular \mbox{$:n_k: =
\delta n_k$}) and $\Gamma(K,p,q)$ is the vertex function encoding
all scattering processes induced by the interaction. We define a
CUT which maps the Hamiltonian (\ref{hamilton}) to an effective
one $H_{\rm eff}$ which conserves the total number of quasi-particles
$Q := \sum_k {\rm sign}(|k|-k_{\rm F}):n_k:$, i.e.\
$ [Q,H_{\rm eff}] = 0 $.
To this end, the flow parameter $\ell$ is used to denote intermediate
Hamiltonians $H(\ell)$. Starting point is $H(\ell=0)=H$; end point
is $H(\ell=\infty)=H_{\rm eff}$. The transformation is defined by
\begin{mathletters}
\label{CUT}
\begin{eqnarray}
\label{flow} 
\partial_\ell H  &=& [\eta(\ell), H(\ell)]\\
\label{choice}
\eta_{i,j}(\ell) &=& \mbox{sign}(Q_{i,i}-Q_{j,j})
H_{i,j}(\ell)
\end{eqnarray}
\end{mathletters}
where (\ref{choice}) 
with $\mbox{sign}(0)=0$
defines the infinitesimal generator $\eta$.
The matrix elements $\eta_{i,j}$ and $H_{i,j}$ are given in
an eigen-basis of $Q$. 
The flow stops for $[Q,H]=0$ since then
$\eta=0$. This represents indeed the fix point at $\ell=\infty$
as can be shown generally under certain assumptions \cite{mielk98,knett00a}.
The choice (\ref{choice})
eliminates all parts of $H$ changing the number of quasi-particles.
The block band diagonality in $H(\ell)$
\cite{mielk98,knett00a} is retained which means here
 that the number of quasi-particles is changed only
by $0, \pm 2, \pm 4$. Even the 3- or more-particle terms 
do not change it by $\pm 6$ or higher. It is very important 
that (\ref{CUT}) is {\em renormalizing} since it suppresses matrix elements
between states with very different energies more strongly than
those between similar energies \cite{wegne94,knett00a}.
Since approximations have to be made to solve (\ref{CUT})
this renormalizing property is essential in order not to
loose important low-energy physics.

Let us assume that we are able to solve Eqs.\ \ref{CUT}
exactly and that the resulting flow converges.
 {\em Then the above defined CUT represents Landau's mapping of
quasi-particles}. The conservation $[Q,H_{\rm eff}]=0$ allows to 
describe all excitations by a {\em finite} number of quasi-particles.
 This is one of the crucial points of the present work.
The conservation of the number of quasi-particles 
ensures that terms 
in $H_{\rm eff}$ consisting of  6, 8, or more fermionic operators
become relevant only if three or more quasi-particles are present
in the system. The terms made from 2 and 4 fermionic operators
constitute the 1-particle and the 2-particle parts of the effective
Hamiltonian. They represent the dispersion and the effective
interaction (cf.~Ref.\onlinecite{wegne94}). 
The basic idea of FL theory is that
at low energies only the 1-particle and the 2-particle parts are relevant
even though 3-particle terms are present. 
Comparing Eq.~\ref{energfunc} and Eq.~\ref{hamilton} at $\ell=\infty$
the vertex function at zero momentum transfer ($p=q$) can be 
identified with Landau's interaction function $f(K-p,K+p)=\Gamma(K,p,p)$.
For non-zero momentum transfer $\Gamma(K,p,q)$ can be interpreted as
general quasi-particle scattering amplitude. A precise one-to-one
correspondence, however, to the quantities known from conventional FL
theory \cite{nozie97} cannot be established since the CUT keeps all
interactions local in time so that the vertex function $\Gamma$ in
Eq.~\ref{hamilton} has no frequency dependence.

So the CUT defined in Eq.~\ref{CUT} allows to compute
Landau's interaction function $f$ directly for a microscopic model.
But there is a caveat. Terms comprising six and more fermionic operators
do not vanish and they influence the 1- and 2-particle terms for 
$0<\ell<\infty$ because the quasi-particle conservation is achieved only
at the end of the transformation ($\ell=\infty$). 
The restriction to the scattering processes labelled by the vertex
function $\Gamma$ constitutes an approximation. It is related to 
a 1-loop RG  calculation for the interaction
vertex or, equivalently, to
the summation of the parquet diagrams \cite{solyo79}. Recently, such 
RG computations are performed by numerous groups 
\cite{zanch97,halbo00a,honer01,binz01} for 2D Hubbard 
models, for which  results by a CUT \cite{grote01}
exist, too. There are three particularly
attractive features of the CUT approach (\ref{CUT}). 
First, it is local in the flow parameter $\ell$. Second,
there is no need to take a retardation into account, i.e.\ no
frequency dependence occurs because
 the unitarity of the transformation ensures that
all intermediate Hamiltonians stay hermitian. 
This is a conceptual advantage over the RG procedures used so far
since the book-keeping related to
another dimension, namely time, is avoided. 
Third, and most importantly, a CUT 
 keeps states a {\em all} energies. No information is lost. 
This major conceptual advancement makes it possible
to address also non-universal behavior like the full frequency
dependence of spectral densities \cite{knett01b}.
\end{multicols}
\begin{figure}[htb]
  \begin{picture}(16.8,3.7)
    \put(0.7,0.1){
      \includegraphics[width=5.2cm]{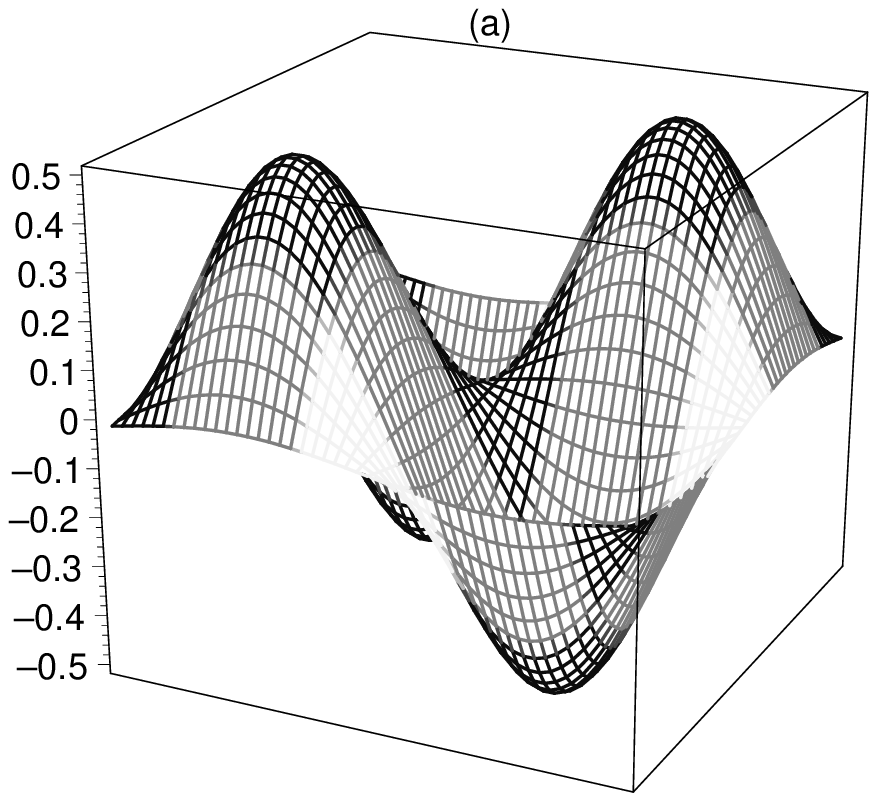}
      \includegraphics[width=5.2cm]{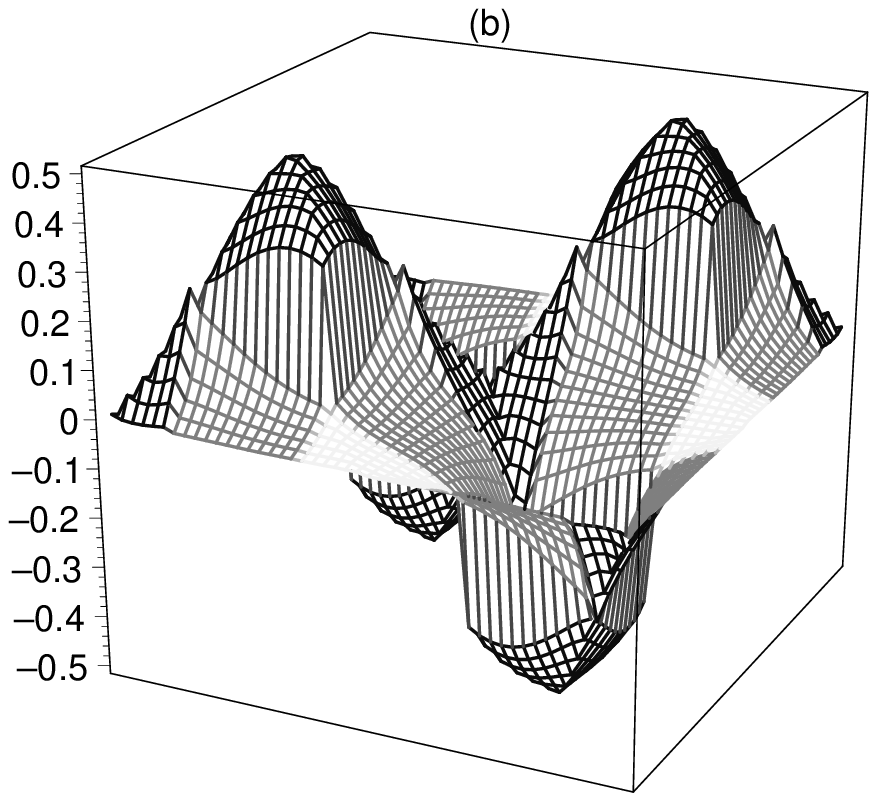}
      \includegraphics[width=5.2cm]{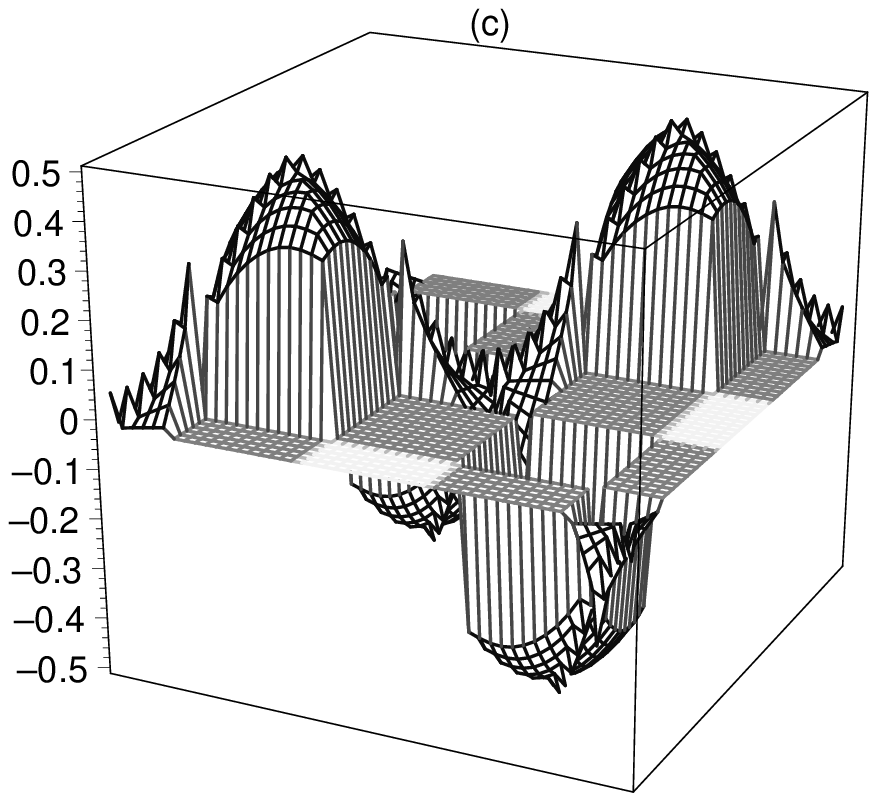}
      }
  \end{picture}
  \caption{Cuts of $\Gamma(K,p,q)$ in Eq.~\ref{hamilton}
    for $K=\pi/4$ and $p, q \in [-\pi,\pi]$: 
    (a) beginning ($\ell=0$), (b) middle ($\ell=0.5$), 
    (c) end of transformation
    ($\ell=\infty$). Black lines stand for processes changing
    the number of quasi-particles by $0$; dark grey for changes by $\pm4$;
    light grey for changes by $\pm 2$.}
\end{figure}

\begin{multicols}{2}
  In order to back up the above general results
  we performed a numerical calculation for a tight-binding model of 
interacting fermions without spin on a chain with average filling
$n=1/2$. The bare dispersion $\varepsilon_0(k)$ reads $-\cos(k)$.
 The interaction is taken to be a nearest-neighbor repulsion
implying $\Gamma_0(K,p,q) := V\sin(p)\sin(q)$ \cite{notiz2}.
This model is equivalent
to an anisotropic, antiferromagnetic Heisenberg spin chain with $S=1/2$
via the Jordan-Wigner transformation \cite{fradk91}. It is exactly
solvable so that much is known about its properties \cite{yang66a,yang66b}.
The system  is metallic  for
$V\le 1$. For larger values the discrete translational invariance is broken
and a gap is opened making the system
insulating; here we do not consider this phase.
The metallic phase itself is a generic realization of a LL
\cite{halda80,halda81a}. The Luttinger parameter can be deduced
from the comparison of exactly known quantities like velocities \cite{johns73}
to the predictions of bosonization \cite{halda80,halda81a}.
So the power-law behavior in the momentum distribution
at the Fermi edge is known quantitatively
\cite{halda81a,voit95}. This power-law behavior and the
absence of a jump in the momentum distribution is
taken as signature for the absence of quasi-particles and the 
prevailance of bosonic modes. Thus the ansatz (\ref{hamilton})
in terms of fermions may seem inappropriate. Yet this is not the case.

We use the parametrization in Eq.~\ref{hamilton} with $\ell$-dependent
energy constant $E_0$,
dispersion $\varepsilon$, and vertex function $\Gamma$. 
In  the limit of an
infinite system the starting values are $\Gamma(\ell=0)=\Gamma_0$,
$\varepsilon(\ell=0)=-(1+2V/\pi)\cos(k)$, and $E_0(\ell=0)=(-(1+V/2)/\pi) N$,
where $N$ is the system size. The deviations from the bare values result
from the normal-ordering assumed in Eq.~\ref{hamilton}. 
The Eqs.~\ref{hamilton} and \ref{CUT} form a closed set of
equations once the normal-ordered 3-particle terms engendered on the
r.h.s.\ of Eq.\ \ref{flow} are omitted \cite{notiz2}.
 We treat the resulting
set of functional differential equations by discretization, i.e.\
we solve the equations for finite systems. Up to about $N=50$ sites 
could be treated well numerically, which amounts up to a set of
several thousand differential equations.
 In Fig.~1, cuts of the vertex
function $\Gamma$ for a given total momentum $K$  illustrate
the transformation at work. Different
grey scales encode the action of the scattering process on the number
of quasi-particles. Clearly, the processes that change the quasi-particle
number are gradually suppressed (light and dark grey)
 on $\ell\to\infty$ so that at $\ell=\infty$
only the processes (black) remain which leave the number of quasi-particles
unchanged. 

\begin{figure}[htb]
 \begin{center}
   \includegraphics[width=8cm]{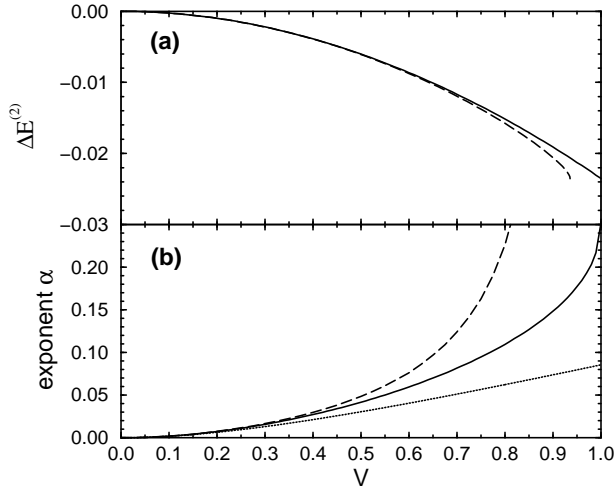}
    \caption{(a) Ground state energy per site beyond absolute and linear term:
    exact for $N=\infty$ (solid); CUT  for $N=48$ (dashed). 
    (b) Momentum distribution exponent 
    $\alpha$: exact for $N=\infty$ (solid);
    CUT  for $N=48$ (dashed),  details in main text; 
    bosonization for $N=\infty$ (dotted).}
\end{center}
\end{figure}
In  Fig.\ 2a the ground state energy beyond the
absolute and linear term in $V$, i.e.\ the correlation part, is
depicted. The approximate CUT result agrees well with the exact result.
Only for $V\gtrapprox 0.75$ a deviation becomes discernible; beyond 
$V_{\rm si}\approx 0.94$ the approximate CUT treatment breaks down because
of the omission of the 3-particle terms. This omission leads 
to a spurious instability (subscript $_{\rm si}$). 
Scaling the interaction $V_{\rm si}$ where the spurious breakdown occurs
to infinite system size we find $V_{\rm si}(N=\infty)\approx 0.75$
which shows that there is a finite range of interactions where the CUT
restricted to the ansatz (\ref{hamilton}) works.

More information on the underlying physics is provided by the
the  momentum distribution $n(k):=\langle n_k \rangle$ which 
can be computed  by  $n(k)=N \partial E/\partial
\varepsilon_0(k)$. 
A generic result is shown in Fig.~3 (crosses). 
At first glance, the data points towards
a fairly large jump. But the points close
to the Fermi surface bend strongly towards the value 1/2.
Similar CUT results were obtained for a Tomonaga model
\cite{hofst00}. For a quantitative
analysis we use a  FL fit 
\cite{noack99}  and a  LL fit \cite{voit95}
 as shown in Eqs.~\ref{fermi} and \ref{luttinger}
\begin{mathletters}
\label{fits}
\begin{eqnarray}
\left|n(k)- 1/2\right| &\approx&
Z/2+ \Delta k\left(B\ln(\Delta k) + C\right)
\label{fermi}
\\
\left|n(k)- 1/2 \right| &\approx&
B \Delta k^\alpha+C \Delta k
\label{luttinger}
\end{eqnarray}
\end{mathletters}
with $\Delta k = |k-k_F|$;
the parameters $Z_{k_F}, B, C$ and $\alpha, B, C$ are fitted.
Checking the applicability of (\ref{luttinger}) on finite-size
data obtained by bosonization for the Luttinger model \cite{meden96}
shows that least-square fits are most robust and provide good
estimates within 5\% accuracy for the exponent $\alpha$.
In Fig.~3, the excellent agreement of
the LL type fit and the  much poorer agreement of the Fermi type fit
are salient. This result is not affected by the precise fit procedure.
\begin{figure}[htb]
  \begin{center}
    \includegraphics[width=7.8cm]{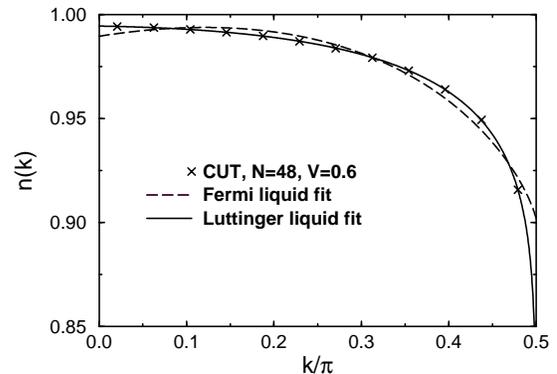}
    \caption{Momentum distributions (data is point-symmetric about 
      $(\pi/2,1/2)$); Fermi liquid fit after Eq.~\ref{fermi}; Luttinger 
      liquid fit with $\alpha= 0.076$ after Eq.~\ref{luttinger}.}
  \end{center}
\end{figure}

By numeric evaluation, we cannot {\em prove} the power law behavior.
For instance, previous results \cite{wegne94}, which rely on additional
approximations in the transformation of the observable, yielded only a
logarithmic correction. We find that our numerical data is better described by
(\ref{luttinger}) than by a  $\ln$-type fit $B+a\ln(\Delta k) +C\Delta k$ just
 in the same way the data in Ref.~\onlinecite{meden96} is better fitted by
(\ref{luttinger}) than by the $\ln$-type fit.

In addition, the exponents $\alpha$ in  Fig.~2b obtained from the
fits agree very well with the exact result.
Thus the excellent agreement of the
LL fit in Fig.~3 is not accidental. For $V < 0.6$ 
the approximated CUT result agrees even better with the exact result than does
the straight application of bosonization. While the bosonization result
is much too smooth the approximated CUT overestimates the exponent due to the
neglect of the 3- and more-particle terms. 

The very good agreements in Figs.~3 and 2b provide evidence 
that the CUT (\ref{hamilton},\ref{CUT})
targeting on quasi-particle conservation captures the relevant
LL behavior. This comes as a surprise since the 
concept of holes below and particles above the Fermi level as
quasi-particles belongs  in the first place to FL theory.
This shows that a powerful RG
procedure such as the CUTs allows one to stick to the
FL type of quasi-particles. An analogous observation
was made previously for  spinons and triplets \cite{knett01b}.
 It
is decisive to describe the multi-particle dynamics properly.

The validity of our mapping onto conserved fermionic
quasi-particles is strongly supported by
 recent results  \cite{kehre99,kehre01} for the closely
related quantum sine-Gordon model. 
By a similar CUT Kehrein  mapped this strongly correlated model,  which
corresponds to an {\em interacting} fermionic model \cite{colem75}, for a wide
range of parameters onto non-interacting fermions \cite{notiz3}.

A description of LL behavior  in terms of fermionic
quasi-particles is intriguing since one can assume that the
dimensional crossover to higher dimensional FL behavior
is also captured: for $D>1$, the quasi-particle description should
work even better. Existing RG approaches provide  good insight
in the dimensional crossover in terms of effective low-energy
models \cite{ueda84,bourb91,caste94}. Renormalizing CUTs yield
a quantitative description on {\em all} energy scales since no
degrees of freedom are integrated out. Besides the advantages of
the locality in the flow parameter $\ell$ and of the avoidance of 
retardation, this feature constitutes the main progress.
In this way,  spectral properties with full frequency dependence 
are accessible \cite{knett01b}.

Summarizing, we presented a  route based on 
continuous unitary transformations (CUT) to the explicit
 construction of Landau's
mapping of the excitations of the interacting system to the
ones of the non-interacting system.
The importance of quasi-particle conservation was highlighted.
We constructed Landau's mapping numerically in the worst
case, namely for  one-dimensional fermions. The results indicate that 
the CUT approach  captures a finger-print of Luttinger
liquids: power-law behavior in the momentum distribution.
This finding suggests the CUT approach as method to 
describe the crossover from Fermi liquids to Luttinger liquids
on all energy scales.

We thank S.~Kehrein, C.~Knetter, V.~Meden, W.~Metzner,
 E.~M\"uller-Hartmann, K.~Sch\"onhammer, and F.~Wegner
for helpful discussions. Financial support of the
DFG in SP1073 is acknowledged.


\begin{thebibliography}{10}
\bibitem{notiz3}
Kehrein uses the generator $\eta$ proposed 
by Wegner \cite{wegne94}. In Refs.~\cite{kehre99,kehre01}, no
quasi-particle interaction is considered although it is
 also generated by the CUT.
\end{thebibliography}





\end{multicols}
\end{document}